\documentclass{emulateapj}

\usepackage{amsmath,amssymb,verbatim} 
\usepackage{color} 
\usepackage{natbib}

\shorttitle{Resistive Pulsar Magnetospheres}
\shortauthors{}

\begin{document}
\title{Resistive Solutions for Pulsar Magnetospheres}


\author{Jason Li$^1$, Anatoly Spitkovsky$^1$, \& Alexander
Tchekhovskoy$^2$} \affil{$^1$Department of Astrophysical Sciences, Peyton Hall,
Princeton University, Princeton, NJ 08544, USA\\ $^2$Princeton Center for
Theoretical Science, Jadwin Hall, Princeton University, Princeton, NJ
08544, USA} 
\email{jgli@astro.princeton.edu}

\begin{abstract}
The current state of the art in the modeling of pulsar magnetospheres
invokes either the vacuum or force-free limits for the magnetospheric
plasma. Neither of these limits can simultaneously account for both
the plasma currents and the accelerating electric fields that are
needed to explain the morphology and spectra of high-energy emission
from pulsars. To better understand the structure of such
magnetospheres, we combine accelerating fields and force-free
solutions by considering models of magnetospheres filled with
resistive plasma.  We formulate Ohm's Law in the minimal velocity
fluid frame and construct a family of resistive solutions that
smoothly bridges the gap between the vacuum and the force-free
magnetosphere solutions. The spin-down luminosity, open field line
potential drop, and the fraction of open field lines all transition
between the vacuum and force-free values as the plasma conductivity
varies from zero to infinity. For fixed inclination angle, we find
that the spin-down luminosity depends linearly on the open field line
potential drop.  We consider the implications of our resistive
solutions for the spin down of intermittent pulsars and sub-pulse
drift phenomena in radio pulsars.

\end{abstract}



\section{Introduction}\label{sec:intro}

Pulsar magnetospheres are filled with plasma, and the presence of
plasma affects all magnetospheric properties. Although this fact has
been well appreciated since the early days of pulsar research, the
ability to quantitatively model plasma effects has emerged mainly in
the last decade.  Generally speaking, the models of pulsar
magnetospheres can be classified according to the amount of plasma
supply they assume. At one extreme is the vacuum magnetosphere, which
is modeled with the magnetic field of a spinning inclined dipole in
vacuum \citep{Deutsch55}.  As this solution has no plasma, it has no
possibility of producing any pulsar-like emission. However, the fact
that this field is described by an analytic formula has made it the
most widely used framework for calculating pulsar properties. For
example, the characteristic spin-down energy loss which yielded the
fiducial pulsar field strength of $10^{12}$G or the polarization sweep
of the rotating-vector model \citep{RadCooke69} are guided by the
vacuum field shape. Slot-gap and outer gap models of gamma-ray
emission from pulsars are also based on this field \citep{CHR86a,
DHR04}.  The next order of approximation is models that allow plasma
emission from the surface of the star. These include the original
charge-separated model of \citet[hereafter GJ]{GJ69}, and the space
charge-limited models with pair formation
\citep[e.g.,][]{rudermansutherland75,AS79}. These models envision both
the regions where plasma shorts out the accelerating electric fields
and the vacuum-like regions where acceleration is present (so-called
``gaps''). This approach allows for more realism in studying plasma
creation and acceleration, but at the price of being decoupled from
the global structure of the magnetosphere, as the modification of the
vacuum field due to plasma currents is typically not included.
Finally, the models that concentrate on the global magnetospheric
properties assume that abundant plasma exists throughout the
magnetosphere and in the wind. This plasma shorts out the accelerating
electric fields and provides the corotation of field lines with the
star. Such models include the relativistic magnetohydrodynamic (MHD)
description of the magnetosphere and its limit for low-inertia
magnetically-dominated plasmas, the ``force-free'' description. Which
of these different regimes is applicable to real pulsars may
ultimately depend on the physics of plasma supply in the
magnetosphere.

Currently, quantitative solutions of the global magnetospheric
structure exist only for the vacuum limit \citep{Deutsch55} and for
the limit of abundant plasma in force-free electrodynamics (see
\citealt{CKF99, Gruzinov05, Timokhin06, McKinney06,
Komissarov06}\footnote{We note that \cite{Komissarov06} has also
calculated the structure of the aligned rotator in the relativistic
MHD limit including particle inertia.}  for aligned rotators, and
\citealt[hereafter S06]{Spitkovsky06}; \citealt{CK10} for pulsars with
arbitrary inclinations).  The real pulsar magnetosphere is likely
operating somewhere in between these limits, with various accelerating
gaps, regions of pair production, and strong current sheets likely
causing local violations of the ideal MHD constraint, $\vec{E}\cdot
\vec{B}=0$. Knowing the structure of the magnetosphere, including such
non-ideal effects, would be very useful for calculating the properties
of pulsar emission. Indeed, currently the ideal force-free models that
include the back-reaction of plasma currents on the field structure
lack any accelerating fields by construction, and thus cannot be used
to directly predict the spectra of gamma-ray radiation observed by
Fermi GST.

One way to reintroduce accelerating electric fields in the
magnetosphere is to allow finite resistivity of the plasma. Several
formulations of resistive force-free equations have been proposed,
most notably by \cite{Lyutikov03} and \cite{Gruzinov07,Gruzinov08}. In
resistive MHD, the Ohm's law can be unambiguously defined in the
proper frame of the fluid by relating the current in that frame to the
electric field through $\vec{j}_{\rm fluid}=\sigma \vec{E}_{\rm
fluid}$, where $\sigma$ is plasma conductivity \citep[see,
e.g.,][]{Lichnerowicz67, Palenzuela09}. In the force-free system,
however, the fluid velocity along the magnetic field is unknown, and
only the transverse velocity can be obtained from the electromagnetic
fields. This introduces some freedom in prescribing the Ohm's law in
the force-free picture. In the prescription proposed by
\cite{Lyutikov03}, the parallel velocity along the field was taken to
be zero. In ``Strong-Field Electrodynamics'' \citep[hereafter
SFE]{Gruzinov07,Gruzinov08}, the Ohm's law was formulated in the frame
that moves along the field lines with such speed that the charge
density in that frame vanishes. Despite the fact that such a frame
formally exists only in space-like regions, SFE prescription appears
to give smooth numerical solutions throughout the magnetosphere
\citep{Gruzinov11}, and, most importantly, does not explode in current
sheets where magnetic fields can go through zero. Both formulations
are arbitrary, however, because the real fluid velocity does not have
to follow either frame assumption. As an additional constraint, it is
useful, therefore, to construct a resistive formulation that
reproduces physical solutions expected at the extremes of very large
and very small conductivity of the plasma, namely the force-free limit
($\sigma\to\infty$) and the limit of vacuum electromagnetism
($\sigma\to 0$). Below we describe a resistive prescription that
generalizes these schemes and combines the correct limiting behavior
of Lyutikov's scheme with current-sheet stability of Gruzinov's
formulation. We then apply this prescription to numerically calculate
the structure of resistive magnetospheres in pulsars.

In the resistive force-free picture of the pulsar magnetosphere, the
magnetized neutron star is thought to be surrounded by an abundant
massless plasma with finite conductivity, so that not all accelerating
fields are shorted out. For simplicity and as a proof of principle, we
only consider the unrealistic case of constant conductivity throughout
space. More complicated prescriptions will be studied elsewhere. Our
finding is that using our formulation of the resistive force-free
electrodynamics we can construct a family of magnetospheres that
smoothly transition from the Deutsch vacuum solution to the ideal
force-free magnetosphere as the conductivity of the plasma is
increased. Such intermediate magnetospheres possess interesting
properties that we discuss in this paper.  We study the variation of
the spin-down power with magnetic inclination angle as a function of
plasma conductivity and relate it to physical conditions such as the
effective potential drop on the open field lines. In
\S\ref{sec:currentder} we discuss the derivation of the resistive
force-free prescription, in \S\ref{sec:simulations} we describe our
code for solving the equations and present sample magnetospheric
solutions. Discussion and potential applications to pulsar physics are
in \S\ref{sec:discussion}.

\section{Resistive Electrodynamics}\label{sec:currentder}
We describe here our prescription for resistive current as used in our
numerical code.  Ohm's Law is defined in the fluid rest frame.  In
this frame the electric and magnetic fields are parallel; otherwise,
there would be a particle drift across magnetic field lines.  The
laboratory frame can be connected to the fluid rest frame through two
boosts.  One boost is in the $\vec{E}\times\vec{B}$ direction in the
lab frame and brings the electric field parallel to the magnetic
field.  The other boost is along the parallel electric and magnetic
fields and transforms the 4-current to the fluid rest frame while
leaving the electric and magnetic fields parallel.  We choose a simple
Ohm's Law to relate the current and electric field in the fluid frame:
$\vec{j}_{\rm fluid}=\sigma \vec{E}_{\rm fluid}$, where $\sigma$ is
the plasma conductivity.  More general formulations of the Ohm's Law
that take into account time-dependent currents, inertial effects,
pressure, and the Hall effect have been derived \citep[see
e.g.,][]{Meier04}. It is likely safe to ignore most of these effects
in the bulk of the strongly magnetized cold pair plasma, as envisioned
for typical pulsar magnetospheres; however, singular current sheets
may require a more elaborate treatment.  In this work we use a
constant uniform conductivity throughout the domain and restrict
ourselves to the simple Ohm's Law to elucidate the basic physics.
Boosting back to the laboratory frame, the current vector can be
expressed as
\begin{eqnarray}
\label{generalcurrent}
\vec{j}&=& \frac{\rho c\vec{E}\times\vec{B}}{B^2+E^2_0} \nonumber \\ 
&+&\frac{(-\beta_{||}\rho c+\sqrt{\frac{B^2+E^2_0}{B^2_0+E^2_0}(1-\beta_{||}^2)}\sigma E_0)(B_0\vec{B}+E_0\vec{E})}{B^2+E^2_0}.
\end{eqnarray}
See Appendix A for a full derivation of this expression and the
subsequent limits that we discuss below. Here, $\beta_{||}$ is the
magnitude of the boost along the parallel electric and magnetic fields
to the fluid rest frame, $\rho$ is the charge density in the
laboratory frame, and $E_0$ and $B_0$ are the magnitudes of the
electric and magnetic fields in all frames in which they are parallel.
$E_0$ and $B_0$ are defined by the expressions
\citep{Gruzinov07,Gruzinov08}
\begin{align}
B_0^2&=\frac{\vec{B}^2-\vec{E}^2+\sqrt{(\vec{B}^2-\vec{E}^2)^2+4(\vec{E}\cdot \vec{B})^2}}{2}, \nonumber \\
E_0&=\sqrt{B_0^2-\vec{B}^2+\vec{E}^2}, \nonumber \\
B_0&={\rm sign}(\vec{E}\cdot\vec{B})\sqrt{B^2_0},
\label{eq:E0andB0}
\end{align}
where we allow $B_0$ to be positive or negative, depending on whether
the magnetic field is aligned or antialigned with the electric field.
Note that in addition to the advection of charge in the
$\vec{E}\times\vec{B}$ direction, there is conduction current along
both the laboratory frame magnetic field and the electric field (the
last term in equation \ref{generalcurrent}).  There is still a
principal ambiguity here in defining the fluid rest frame, as we have
not yet specified the magnitude of the parallel boost $\beta_{||}$.  In
fact, this speed cannot be obtained from a purely electrodynamic
standpoint without the inclusion of gas dynamics, so a suitable choice 
has to be made. 

One particularly interesting choice of fluid frame is the slowest
moving frame that has electric fields parallel to magnetic fields,
i.e., the frame with $\beta_{||} \to 0$.  If particles start out on
the stellar surface with non-relativistic velocity, then the fluid essentially
satisfies this condition in the high conductivity limit \citep[see
e.g.,][]{ContopKazan02,Tchekho09,Beskin10}.  In this minimal velocity
limit the current becomes
\begin{equation}
\vec{j}=\frac{\rho c\vec{E}\times\vec{B}+\sqrt{\frac{B^2+E^2_0}{B^2_0+E^2_0}}\sigma E_0(B_0\vec{B}+E_0\vec{E})}{B^2+E^2_0}.
\label{current}
\end{equation}
This form of the current has the
especially useful property that in the limit of vanishing conductivity
the conduction current along the electric and magnetic fields goes to
zero.  Such a magnetosphere does not develop space charge due to the
flow of charge from the star along the field lines.  Thus, the current
in the $\vec{E}\times\vec{B}$ direction also vanishes.  This is
exactly what we expect in the vacuum limit.  Since we are interested
in producing a resistive transition from ideal force-free to vacuum,
we use this form of the current in our numerical investigation.
\cite{Lyutikov03} derived an expression for resistive current in
magnetically dominated plasma, choosing the same minimal velocity
frame for the formulation of the Ohm's law as in our
derivation. However, his boost in the $\vec{E}\times\vec{B}$ direction
to the fluid frame contains an error and does not bring the electric
and magnetic fields parallel to one another. As a result, our final
expressions differ.

We can alternatively express the minimal velocity current as
\begin{equation}
\vec{j}=\rho \vec{v}+\sigma \vec{E}_{\rm fluid}, 
\end{equation}
where the fluid velocity $\vec{v}=c(\vec{E}\times\vec{B})/(B^2+E_0^2)$
is the generalized drift velocity, $\vec{E}_{\rm
fluid}=\gamma(\vec{E}+\vec{v}\times\vec{B})$, and
$\gamma=(1-v^2/c^2)^{-1/2}$.  The drift velocity contains the
term $E_0^2$ in the denominator to account for the non-zero electric
field in the fluid frame (see equation \ref{eq:betax}).  The presence
of this term allows the current to remain nonsingular in current
sheets, where the magnetic field vanishes.

It is instructive to consider other limits for the current.
\cite{Gruzinov07,Gruzinov08} derived an alternate limit of the general
current in equation (\ref{generalcurrent}), known as SFE.  He
postulated that Ohm's law should be formulated in the frame where
charge density vanishes, which gives parallel boost
\begin{equation}
\beta_{||}=\frac{-\rho}{(\gamma^2_x\sigma^2E^2_0/c^2+\rho^2)^{1/2}}
\end{equation}
and current
\begin{equation}
\vec{j}=\frac{\rho c\vec{E}\times\vec{B}+(\gamma^2_x\sigma^2E^2_0+\rho^2c^2)^{1/2}(B_0\vec{B}+E_0\vec{E})}{B^2+E^2_0},
\end{equation}
where $\gamma_x^2\equiv (B^2+E^2_0)/(B^2_0+E^2_0)$.  We see no special
reason as to why the fluid frame charge density must vanish.  Further,
this current does not reduce to vacuum as the conductivity drops to
zero, as it does in equation (\ref{current}). Consequently, even for
vanishing conductivity the SFE solutions resemble ideal non-resistive
solutions.

In the limit of infinitely conductive strongly magnetized plasma,
which we refer to as the ideal force-free limit, the plasma satisfies
transverse force balance, $\rho\vec{E}+\vec{j}\times\vec{B}=0$, and
the parallel electric field is shorted out, $E_0\to 0$.  However, as
$\sigma\to\infty$, the product of $\sigma E_0$ stays finite, and the
minimal velocity current becomes
\citep{Osherovich88,Gruzinov99,Blandford02}
\begin{equation}
\vec{j}=\frac{\rho c\vec{E}\times\vec{B}}{B^2}+\frac{c}{4\pi}\frac{(\vec{B}\cdot\nabla\times\vec{B}-\vec{E}\cdot\nabla\times\vec{E})\vec{B}}{B^2}. \label{ffreecur}
\end{equation}

\section{Numerical Simulations}\label{sec:simulations}
In order to test the influence of global resistivity on the structure
of pulsar magnetospheres, we employ a three-dimensional numerical code
(see S06) that implements the finite difference time-domain scheme
\citep[FDTD,][]{TafloveHagness05} to evolve electromagnetic fields
from Maxwell's equations,
\begin{align}
\label{maxwell}
\partial_t \vec{E} &=c \vec{\nabla}\times \vec{B} - 4\pi\vec{j},  \nonumber \\
\partial_t \vec{B} &= -c\vec{\nabla}\times \vec{E},
\end{align}
where the current is given by equations (\ref{eq:E0andB0}) and
(\ref{current}).  Electric and magnetic fields are decentered on the
Cartesian Yee mesh, as in standard FDTD. Instead of the leapfrog time
integration typically used in FDTD, we employ a third order
Runge-Kutta integrator as in S06. This is done to remove the
interpolation in time that would otherwise be needed to bring the
$\vec{E}$ and $\vec{B}$ fields to the same time step when computing
the source current.  We run our simulations on a grid of size $1024^3$
at Courant number 0.5.  The central region of our grid is occupied by
a conducting spherical star of radius $R_*$, rotating at angular
velocity $\vec{\Omega}$, with embedded dipole field of magnetic moment
$\vec{\mu}$ inclined relative to the rotation axis by angle $\alpha$
(see S06 for more detail).  The electric field inside the star is set
to ensure rigid corotation, and we smooth fields across the stellar
boundary as in S06 in order to minimize stair-stepping of a sphere on
a Cartesian grid.  The spatial resolution of our simulations is such
that the light cylinder (LC) radius, $R_{\rm LC}=c/\Omega$, is resolved
with 80 cells, and the simulation box is $12.8\, R_{\rm LC}$ on a
side.  The outer boundary condition is periodic, which limits our run
time to the light travel time across the grid.  We set $R_*=30$ cells,
allowing us to resolve the star while keeping the star sufficiently
small compared to the light cylinder.  We run our code for a range of
different values of the dimensionless conductivity, $\sigma /\Omega$,
to produce a transition from vacuum at low conductivity to ideal
force-free solution at high conductivity.  We have verified that our
solutions are converged with spatial resolution, as well as run
sufficiently long so as to reach a steady state in the frame
corotating with the pulsar.

In addition to the resistive solutions obtained with the above method,
we also consider two special cases of the current prescription that we
call ``force-free" and ``vacuum." By force-free solutions we mean
``ideal force-free" magnetospheres; however, instead of using the full
equation (\ref{ffreecur}) for the current, our numerical
implementation follows the approach in S06, where at each time step we
evolve the perpendicular part of the current from equation
(\ref{ffreecur}) and then correct the electric field by removing any
electric field component parallel to the magnetic field. Further field
limiters are used to enforce $E<B$ in current sheets. As a result, the
force-free scheme of S06 has an effective small resistivity, because
the cleaning of parallel fields can be interpreted as the effect of
very large conductivity along field lines. Also, as all parallel
fields are cleaned on every step, such a scheme does not have the
usual convergence behavior with timestep. The resistive scheme using
equation (\ref{current}) gives us a more well-defined method for
handling magnetospheric dissipation. In addition, this resistive
scheme does not require special limiters to handle current
sheets\footnote{This property is not unique to our resistive current
prescription, and is also present in the original formulation by
\cite{Gruzinov07, Gruzinov08}.}.  The ``vacuum" solutions described
below evolve Maxwell's equations without volume charges or currents
outside the star, allowing for significant potential drops along field
lines.

\begin{figure*}[htp]
\centering
\includegraphics[width=0.5\textwidth]{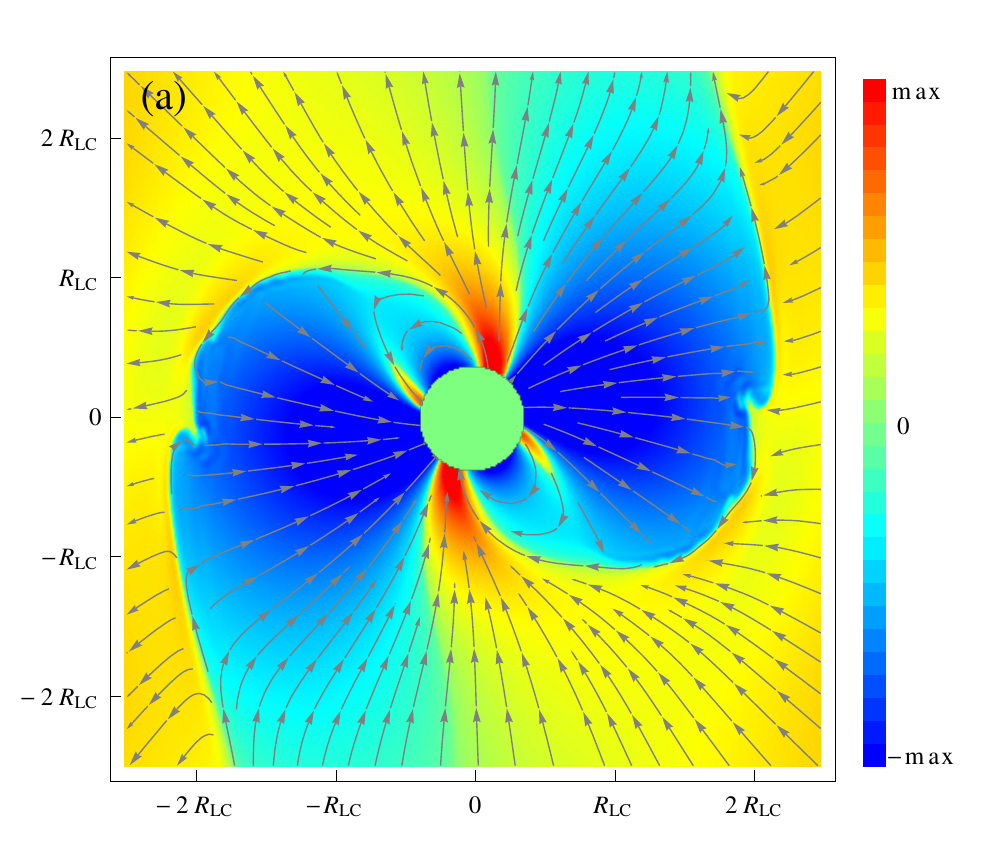}\hfill
\includegraphics[width=0.5\textwidth]{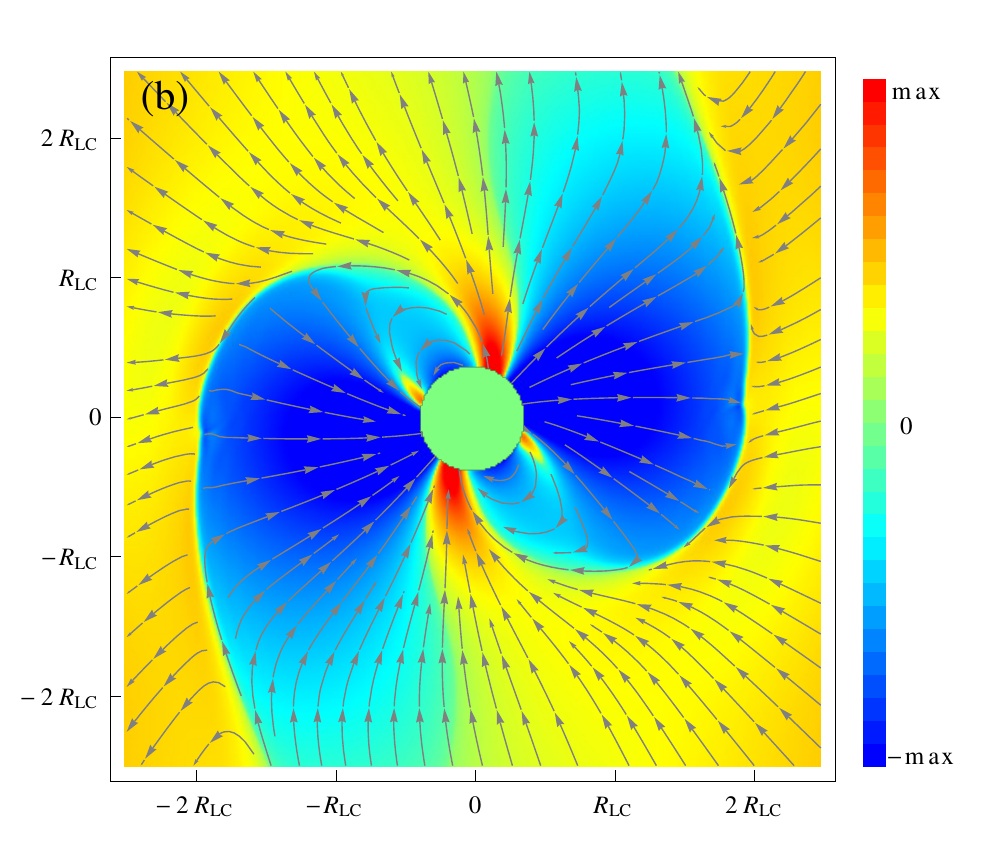}\\
\includegraphics[width=0.5\textwidth]{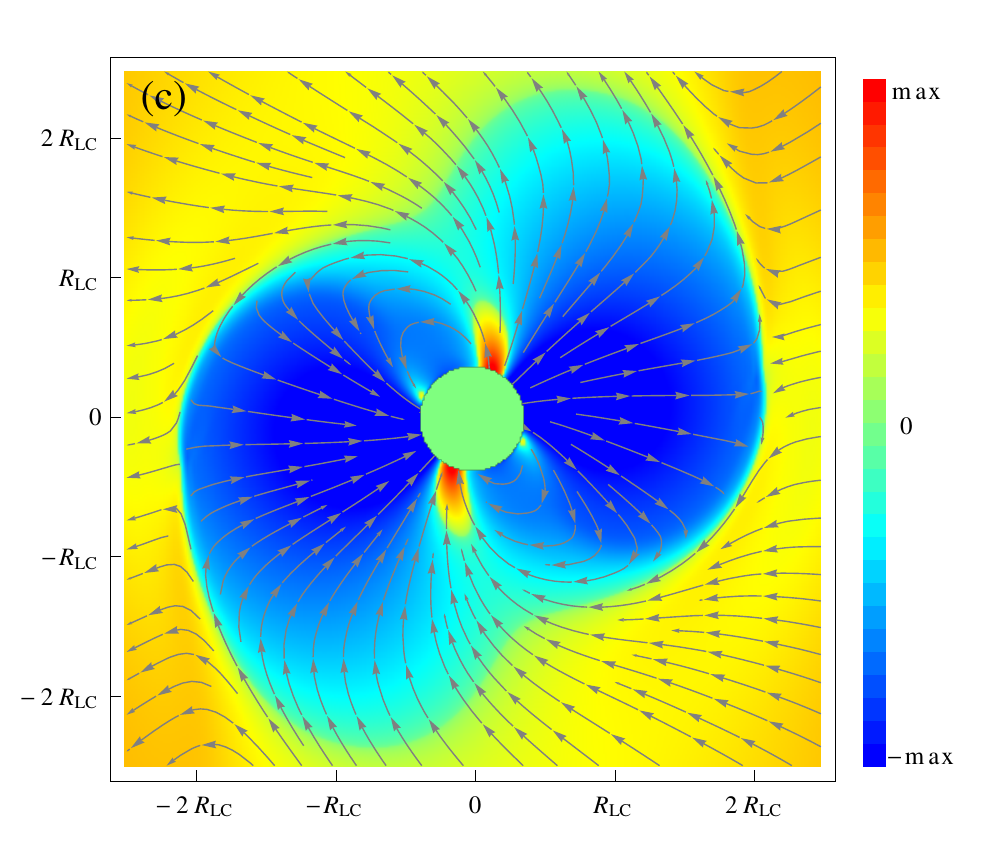}\hfill
\includegraphics[width=0.5\textwidth]{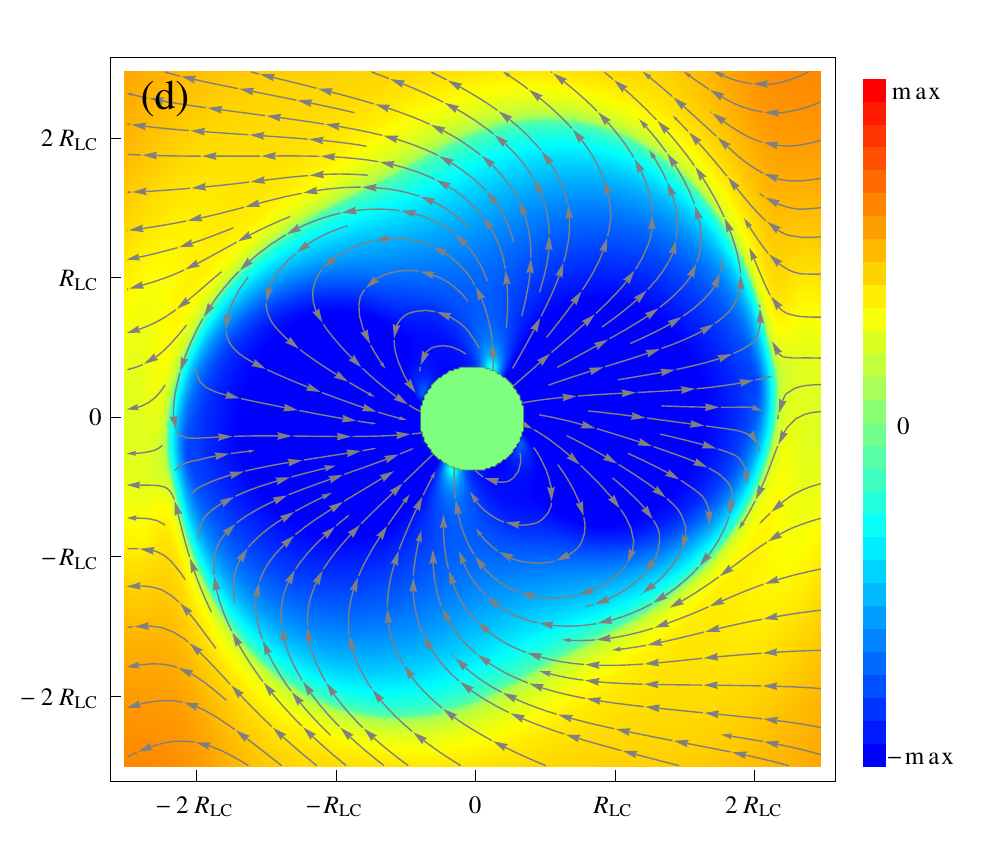}\\
\includegraphics[width=0.5\textwidth]{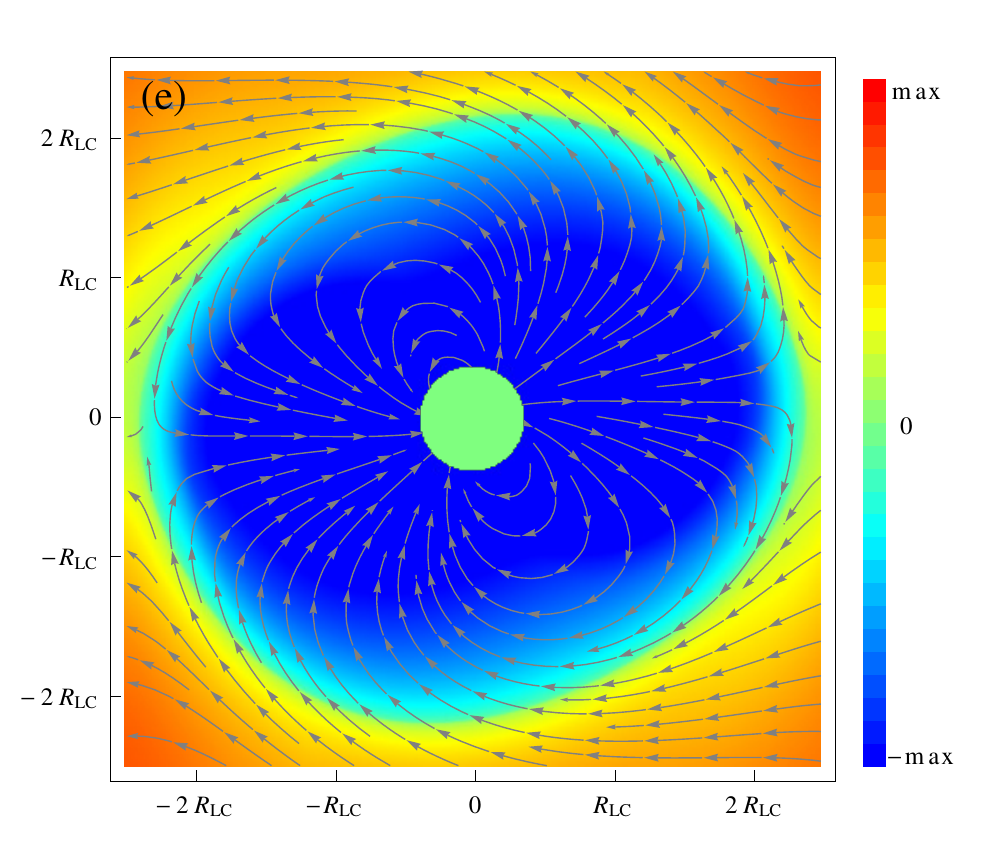}\hfill
\includegraphics[width=0.5\textwidth]{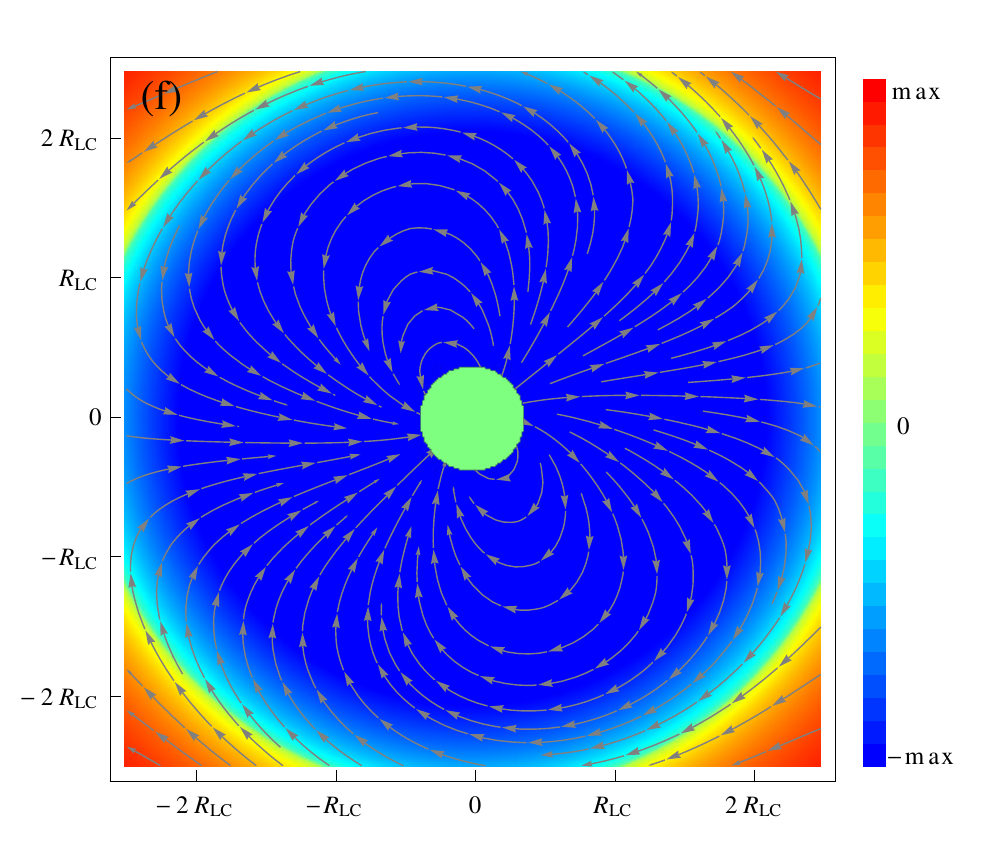}
\caption{Magnetic field lines in the $\vec{\mu}-\vec{\Omega}$ plane
for a $60^{\circ}$ inclined dipole.  Color represents out-of-plane
magnetic field into (red) and out of (blue) the page.  The color table
shows only values up to 30\% of the maximum of the out-of-plane
magnetic field, and a square root stretching has been applied to its
magnitude.  The maximum out-of-plane magnetic field values are given
in Table \ref{maxvals}.  Conduction and displacement currents weaken
with decreasing conductivity.  (a) force-free dipole; (b) resistive
dipole at $(\sigma/\Omega)^2=40$; (c) $(\sigma/\Omega)^2=4$; (d)
$(\sigma/\Omega)^2=0.4$; (e) $(\sigma/\Omega)^2=0.04$; (f) vacuum
dipole.}\label{fields}
\end{figure*}

In Fig.~\ref{fields} we show magnetic field lines in the
$\vec{\mu}$-$\vec{\Omega}$ plane for $60^{\circ}$ inclined force-free
dipole (panel a), resistive dipoles at $(\sigma/\Omega)^2=40$ (panel
b), $(\sigma/\Omega)^2=4$ (panel c), $(\sigma/\Omega)^2=0.4$ (panel
d), $(\sigma/\Omega)^2=0.04$ (panel e), and vacuum dipole (panel f).
The field lines are shown after 1.5 turns, sufficiently long to reach
a steady state out to several light cylinder radii in the corotating
frame.  We show only the central five light cylinders of our
simulation.  Color represents the out-of-plane magnetic field into
(red) and out of (blue) the page.  The maximum magnitude of the
out-of-plane magnetic field value is different in each panel, with
maximum value increasing with conductivity.  Table \ref{maxvals} gives
the maximum values normalized to the maximum vacuum value\footnote{The
discrepancy between maximum field for vacuum and $\sigma=0$ solutions
is numerical and is due to the boundary conditions on the star. This
is further discussed in Appendix B}.  The maximum values occur at the
stellar surface and fall on the blue end of the color table.  To
improve contrast, the color table shows values up to 30\% of the
maximum, and a square root stretching is applied to the data.
Gradients in color reflect the sum of in-plane components of
conduction and displacement currents.

Fig.~1a shows the force-free magnetosphere.  The gross features of
this solution are the same as were discussed in S06\footnote{In
comparing to previous force-free results, we note that
S06 rescaled the total magnetic field when displaying
the out-of-plane component, whereas we rescale only the strength of
the out-of-plane component.}, \cite{CK10} and \cite{BS10}.  The polar
cap consists of the footpoints of open field lines that extend out to
infinity.  Field lines with both ends attached to the star form the
closed field line region, which extends out to a cylindrical distance
of $1 R_{LC}$.  A large scale conduction current flows outwards from
the polar cap along open field lines and returns through the current
sheet, current layer, and a fraction of the neighboring open field
lines.  The current sheets appear as the sharp transition in the
out-of-plane magnetic field, starting at the Y-point at the light
cylinder.  The current layer flows along the boundary of the closed
field line region.  The conduction current is closed in a circuit by a
surface current flowing across the polar cap, a point we will return
to when discussing pulsar spin-down.

At $(\sigma/\Omega)^2=40$, shown in Fig.~1b, the qualitative picture
of the magnetosphere looks quite similar to force-free.  We find that
values $(\sigma/\Omega)^2>40$ give fairly good estimates of highly
conducting magnetospheres for $R_*/R_{\rm LC}=3/8$.  The dimensionless
quantity $\sigma/\Omega=(c/\Omega)/(c/\sigma)$ is the relativistic
analog of the Elsasser number, defined as the ratio of the Alfv\'{e}n
radius in a rotating system, $v_A/\Omega$, to the resistive diffusion
scale \citep{Elsasser46}.  The above threshold corresponds to
resistive diffusion length scale $c/\sigma$ small compared to the
light cylinder, $c/\sigma < 0.16 R_{\rm LC} \ll R_{\rm LC}$, while
satisfying $c/\sigma \lesssim R_{\rm pc}=R_*(R_*/R_{\rm LC})^{1/2}$.
We have also run simulations with our current prescription at higher
conductivity, $(\sigma/\Omega)^2=400$, and find spin-down luminosities
that match the force-free values quite well.  Our explicit scheme
requires very small time step at such high conductivities though, and
it is perhaps more appropriate to use an implicit-explicit
\citep{Palenzuela09} or fully implicit scheme to explore this
regime. We note that the kink in the current sheet seen in the
force-free solution near $2R_{{\rm LC}}$ disappears for resistive
solutions even at high conductivities. It is likely then that this
kink is a numerical artifact of the cleaning procedure used in our
force-free formulation.

\begin{deluxetable}{ccc}
\tablewidth{0pt}
\tablecaption{Maximum out-of-plane magnetic field values, normalized to the maximum vacuum value, for each panel from fig \ref{fields}.}
\tablehead{\colhead{$(\sigma/\Omega)^2$} & \colhead{$B_{\rm out,max}/B_{\rm out,max,vac}$}}
\startdata
force-free & 8.7 \\ \hline
40. & 8.3 \\ \hline
4. & 6.3 \\ \hline
0.4 & 4.0 \\ \hline
0.04 & 2.5 \\ \hline
0  & 1.2 \\ \hline
vacuum & 1. 
\enddata
\label{maxvals}
\end{deluxetable}

Displacement currents play an important role in the highly conducting
magnetospheres of Figs.~1a and 1b.  The displacement currents in the
current sheets actually dominate over conduction currents for the
$60^{\circ}$ rotator.  Displacement currents are not present in the
magnetospheres of aligned rotators, but the strength of the
displacement current in the current sheets of highly conducting
magnetospheres increases monotonically with increasing inclination
angle.  This is in contrast to the conduction currents, whose strength
is independent of inclination angle for fixed conductivity.  Both
displacement and conduction currents weaken considerably with
decreasing conductivity, reflected by smaller gradients in the
out-of-plane magnetic field in panels (b) through (e) of
Fig.~\ref{fields}.  Recall that the maximum in the color table
corresponds to weaker out-of-plane magnetic field at lower
conductivities.  The extent of the closed field line region also
expands with decreasing conductivity, evident in the same sequence of
panels.  We further expect the current sheet to thicken with
decreasing conductivity.  At $(\sigma/\Omega)^2=0.04$, shown in
Fig.~1e, the current sheet should have characteristic width $c/\sigma$
of order $5R_{LC}$. 

There is an obvious increase in the radius of toroidal field sign
reversal beyond the light cylinder as one transitions from the
force-free to the vacuum magnetosphere. This can be understood from
the increase in the winding radius of the spiral pattern for lower
conductivities. The winding of the spiral is determined by the
characteristic speed of the radial outflow multiplied by the pulsar
period. In the force-free case, the radial outflow is given by the
radial component of the $\vec{E}\times \vec{B}$ velocity, which
increases with radius so that the Lorentz factor grows linearly with
cylindrical radius: $\gamma_{E\times B}=\sqrt{1+(R/R_{\rm LC})^2}$
\citep{ContopKazan02}. Thus, the force-free outflow approaches the
speed of light only asymptotically. In the vacuum case, the ``outflow"
is always at the speed of light, so we expect the sign transition of
the field to occur near the half-wavelength of the vacuum wave, or
$\pi R_{\rm LC}$, as is confirmed by Fig.~1f. The force-free case of
Fig.~1a reverses near $2 R_{\rm LC}$, indicative of the smaller
outflow speed near the light cylinder.

\begin{figure}[t]
\centering \includegraphics[scale=.4]{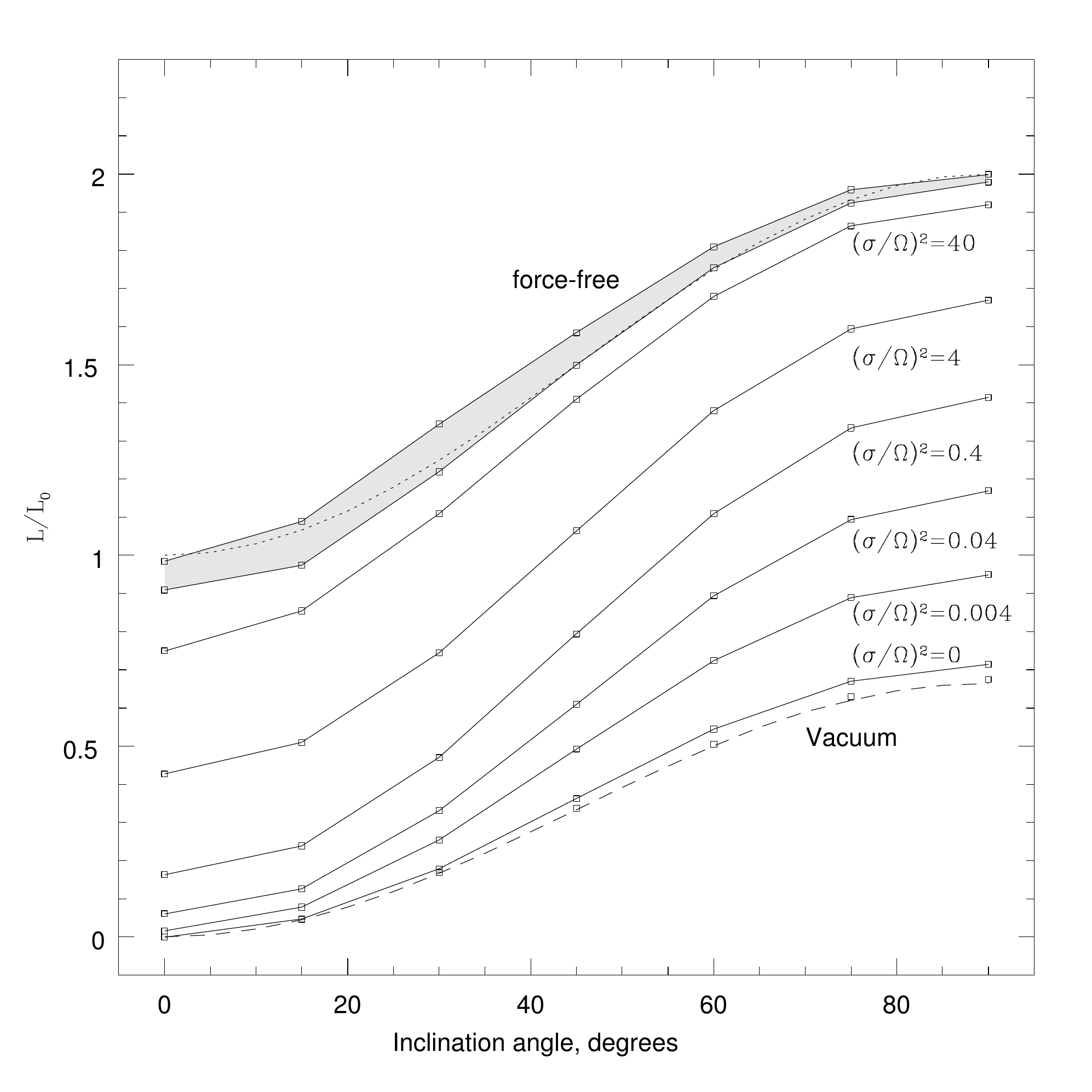}
\caption{Spin-down luminosity dependence on inclination angle for
force-free, a sequence of resistive, and vacuum dipoles.  Spin-down is
normalized by $3/2$ times the spin-down power of the orthogonal vacuum
rotator.  We see a smooth monotonic transition from force-free to
vacuum with decreasing conductivity.}\label{dipole}
\end{figure}

The surface currents flowing across the polar cap of the pulsar exert
a spin-down torque on the star.  The rotational energy loss is
connected to infinity by an outward directed Poynting flux, which can
be thought of as the ejection of toroidal field.  We compute the
Poynting flux at the light cylinder for each of our simulation runs
and show the results in Fig.~\ref{dipole}.  All spin-down values are
calculated for $R_*=3/8R_{LC}$.  The spin-down curves have been
normalized to $L_0$, defined as $3/2$ times the power of the
orthogonal vacuum rotator with finite $R_*=3/8R_{LC}$.
Two-dimensional axisymmetric calculations for small star tell us that
the spin-down of the aligned force-free rotator should be $3/2$ times
larger than the power of the orthogonal point vacuum dipole,
$L_1=2\mu^2\Omega^4/3 c^3$ (\citealt{CKF99,Gruzinov05,McKinney06};
S06).  The short dashed line shows the function
$L/L_0=1+\sin^2\alpha,$ the expected force-free curve in the limit of
small star with inclined dipole (S06). The gray band around the
force-free curve indicates the uncertainty in the measurement due to
boundary effects and numerical dissipation of Poynting flux in the
magnetosphere (see Appendix B for thorough discussion).  Similar bands
are implied but not shown for resistive spin-down curves. The long
dashed line shows the analytic vacuum Deutsch field solution
\citep{MichelLi99}.  One of our principal results is that we see a
smooth monotonic transition from the force-free spin-down curve to the
vacuum spin-down curve for decreasing conductivity.

All spin-down curves in Fig.~\ref{dipole} show a strong dependence on
both inclination angle and conductivity.  The radial Poynting flux
carrying the spin-down power is proportional to the product of
poloidal and toroidal magnetic fields ($E_\theta B_\phi \sim \Omega R
B_p B_\phi /c$), both of which are affected by the strength of
displacement and conduction currents in the magnetosphere. One can get
a qualitative feel for the relative significance of two contributions
by considering the limiting cases. In vacuum, the increase of
spin-down with inclination is solely due to rising displacement
current. This current is likely responsible for much of the angular
dependence of the resistive and force-free solutions. The increase in
spin-down with increasing conductivity is due to the additional
sweep-back and opening of the poloidal field brought on by the
increasing conduction current. We find that spin-down luminosity at
any conductivity has an angle-dependent term proportional to
$\sin^2\alpha$.  We parameterize the spin-down curves in
Fig.~\ref{dipole} with functions $f(\sigma/\Omega)$ and
$g(\sigma/\Omega)$ such that
$L/L_0=f(\sigma/\Omega)+g(\sigma/\Omega)\sin^2\alpha$.  We find the
piecewise linear fit
\begin{align}
\label{eq:spindownsigma}
\frac{L}{L_0}&=0.3+0.3\log(\sigma/\Omega)^2+1.2\sin^2\alpha\,, \,\, (\sigma/\Omega)^2>0.4; \nonumber \\
\frac{L}{L_0}&=0.2+0.08\log(\sigma/\Omega)^2+(1.3+0.2\log(\sigma/\Omega)^2)\sin^2\alpha, \nonumber \\
&0.004<(\sigma/\Omega)^2<0.4.
\end{align}
The amplitude of the angular dependence, $g(\sigma/\Omega)$, is constant for
$(\sigma/\Omega)^2>0.4$ and begins to transition to the vacuum value
below $(\sigma/\Omega)^2=0.4$.

We have thus far shown how spin-down luminosity depends on plasma
conductivity, but the physical meaning of the conductivity is not
entirely clear.  It is instructive to reinterpret the conductivity
parameter, $\sigma/\Omega$, in terms of the potential drop along open
field lines in the corotating frame.  This gives us a handle on the
deviation of the magnetosphere from ideal force-free, which has
vanishing potential drops along field lines.  The electromagnetic
fields in the frame corotating with the pulsar are obtained via a
coordinate transformation from the laboratory frame
\citep{Schiff39,Gron84}:
\begin{equation}
\vec{E'}=\vec{E}+\frac{\vec{\Omega}\times\vec{r}}{c}\times\vec{B}
\label{eq:corotEfield}
\end{equation}
and
\begin{equation}
\vec{B'}=\vec{B}.
\label{eq:corotBfield}
\end{equation}
Since the fields are steady in the corotating frame, $\nabla\times
\vec{E'}=0$ and the corotating electric field can be written as the
gradient of a scalar potential, i.e., $\vec{E'}=\nabla \chi$.  Taking
the line integral of the corotating electric field along a magnetic
field line, $l$, we find
\begin{equation}
\Delta \chi=\int_{l} \vec{E'}\cdot \vec{dl}=\int_{l}
\vec{E}\cdot \vec{dl} \equiv V_{\rm drop}.
\end{equation}
We see that the potential drop along field lines in the corotating
frame can be computed directly from the laboratory frame
fields. Although in resistive solutions particles will actually drift
across magnetic field lines, in addition to accelerating along them,
we choose to study the field-aligned potential drop $V_{\rm drop}$ as
a fiducial measure of particle energy gain.  In fact, the exact
particle trajectories we choose make little difference when estimating
potential drops because the electric field is potential in the frame
corotating with the pulsar.

\begin{figure}[t]
\centering
\includegraphics[scale=.4]{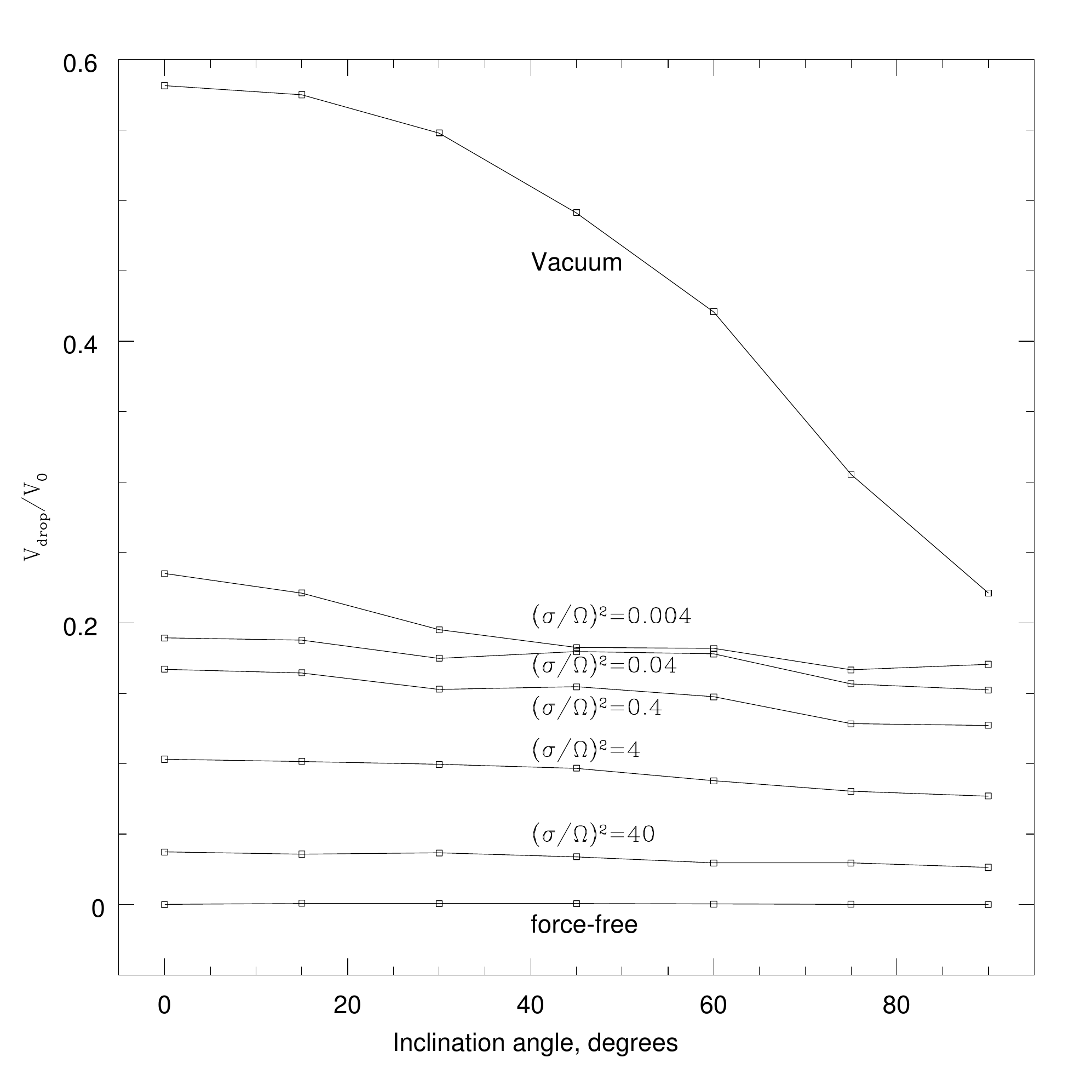}
\caption{Open field line potential drop with inclination angle for
force-free, a sequence of resistive, and vacuum dipoles.  Results are
normalized to $V_0=|\vec{\mu}|/R_{\rm LC}R_*$.  For
$(\sigma/\Omega)^2>0.04$ the potential drop is relatively independent
of inclination angle.  }\label{potential}
\end{figure}

Consider field lines starting on the stellar surface in the
$\vec{\mu}-\vec{\Omega}$ plane separated by $15^{\circ}$ in latitude
from pole to equator.  For every such field line we integrate the
electric field to find the maximum potential drop along each field
line. We then determine the field line with the largest overall
potential drop for a given magnetic inclination.  Integrating field
lines separated by $15^{\circ}$ increments on the stellar surface is
sufficient to give a good estimate of the maximum potential drop along
field lines.  Fig.~\ref{potential} shows the maximal potential drop as
a function of dipole inclination angle for different conductivities.
All results have been normalized to the potential drop from the pole
to the equator of an aligned vacuum rotator in the laboratory frame,
$V_0=|\vec{\mu}|/(R_{\rm LC}R_*)$.  As our models do not prescribe a
high conductivity to the closed field line region, the available
accelerating potential is generally limited by the pole-to-equator
potential difference, rather than the smaller polar cap potential,
$V_{\rm pc}=V_0 R_*/R_{LC}$, with a few notable exceptions.  Low
conductivity solutions, $(\sigma/\Omega)^2 \lesssim 0.04$, at high
inclination angle, $\alpha > 45^{\circ}$, have $V_{\rm drop}$ scaling
intermediate between $V_0$ and $V_{\rm pc}$.  The orthogonal rotator
drop at $(\sigma/\Omega)^2 = 0.04$ scales roughly with $V_{\rm pc}$.
Using the vacuum Deutsch fields, one can obtain that the aligned
vacuum rotator potential drop scales exactly proportional to $V_0$,
whereas the orthogonal vacuum rotator drop scales closer to $V_{\rm
pc}$.  From Fig.~\ref{potential} we see that for
$(\sigma/\Omega)^2>0.04$ the potential drop is roughly independent of
inclination angle for fixed $\sigma$, i.e., there is a one-to-one map
between conductivity and maximal potential drop along a field line
originating in the $\vec{\mu}-\vec{\Omega}$ plane for all inclination
angles.  The maximal potential drop increases from zero for force-free
(or $\sigma/\Omega \to \infty$) to $V_{\rm drop}/V_0\sim 0.2$ for
$(\sigma/\Omega)^2=0.04$.

\begin{figure}[t]
\centering
\includegraphics[scale=.4]{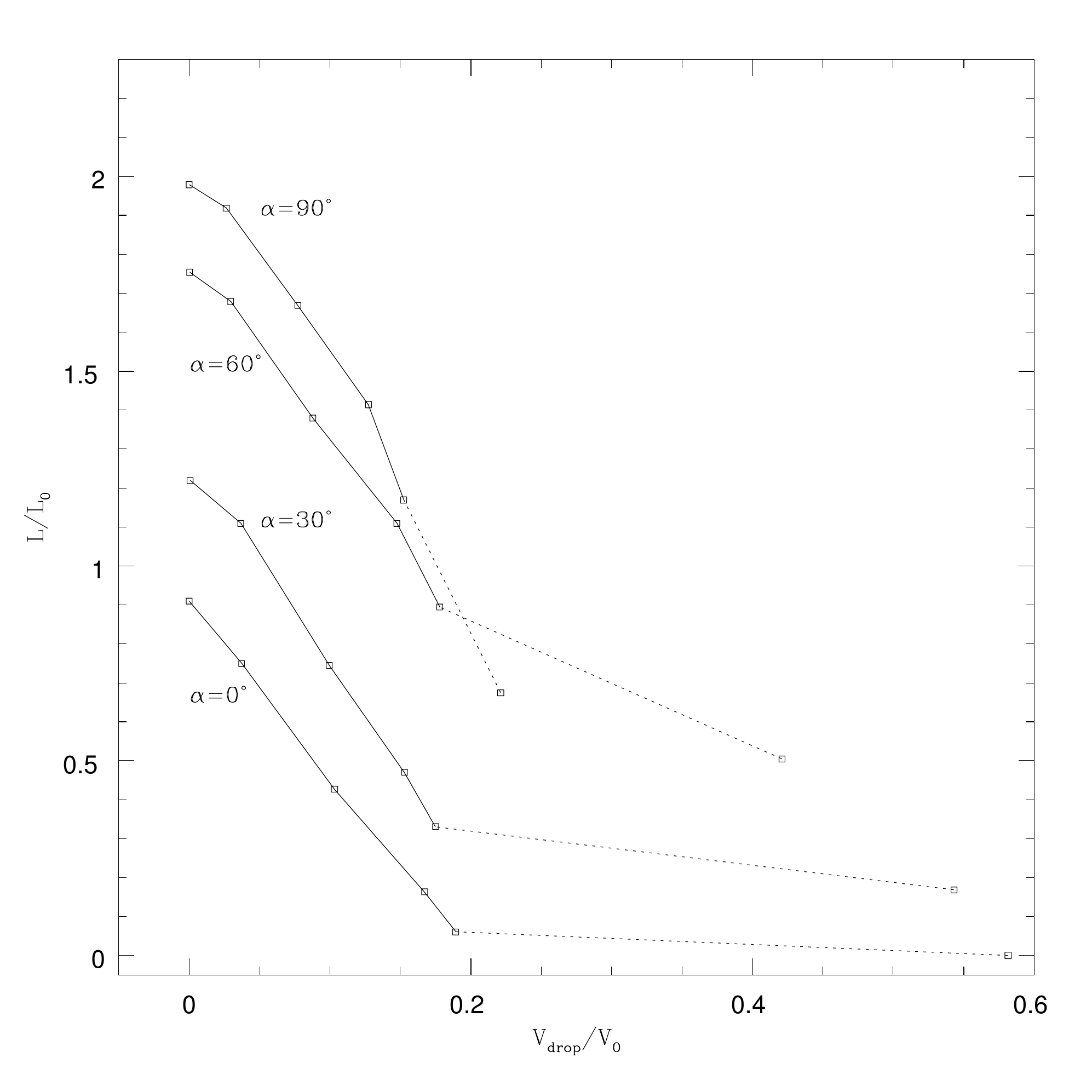}
\caption{Spin down dependence on the open field line potential drop
for a sequence of different inclination angles, $\alpha = 0^\circ,
30^\circ, 60^\circ$, and $90^\circ$.  Potential drop is normalized to
$V_0=|\vec{\mu}|/R_{\rm LC}R_*$ and luminosity to $3/2$ the spin-down
of the orthogonal vacuum rotator.  For $V_{\rm drop}/V_0<0.2$
spin-down falls linearly with potential
drop.}\label{spindown_potential}
\end{figure}

The fact that potential drop is independent of inclination angle for
$(\sigma/\Omega)^2>0.04$ allows us to relate luminosity to potential
drop in a very simple manner.  We present the result here and explain
its origin in more detail below.  Fig.~\ref{spindown_potential} shows
the spin-down luminosity as a function of potential drop for
inclination angles $\alpha=0,30,60,90^{\circ}$.  Spin-down luminosity
increases with increasing inclination angle and with decreasing field
line potential drops.  We fit the spin-down curves in
Fig.~\ref{spindown_potential} with the linear relation
\begin{equation}
\label{eq:spindown}
\frac{L}{L_0}=0.9 \left(1-\frac{V_{\rm drop}}{0.2 V_0}\right)+1.1\sin^2\alpha\,, \,\, 0<V_{\rm drop}< 0.2 V_0.
\end{equation}
This formula applies only to the domain for which we have data shown
with solid lines in Fig.~\ref{spindown_potential}, i.e., $0<V_{\rm
drop}/V_0<0.2$.  A similar spin-down formula was constructed by
\cite{ContopSpitkovsky06} for the case of a finite gap at the base of
the open field lines in a force-free magnetosphere.  The spin-down was
also found to be linear in the potential drop in the gap but had a
different slope and angular dependence of the form $(1-{V_{\rm
drop}/V_{\rm pc}}) \cos^2{\alpha} + \sin^2{\alpha}$. This angular
dependence was not a rigorous derivation, but a prediction based on
aligned force-free and orthogonal vacuum limits.  We note that our
spin-down formula reduces to the force-free spin-down formula from S06
within error bars when $V_{\rm drop}/V_0=0$.

For $V_{\rm drop}/V_0>0.2$ the dependence of spin-down on the
potential drop enters a different regime (see dotted lines in
Fig.~\ref{spindown_potential}).  We do not provide a fit here, as
potential drop no longer scales with $V_0$ at high inclination angle.
The potential drop along field lines in this regime is due to the
reemergence of the vacuum electric fields as conductivity is reduced.
There are three contributions to the vacuum electric field which are
provided by the quadrupolar surface charge on a finite-radius star and
by central and surface monopolar charges. The central monopolar
component provides part of the radial electric field needed for
corotation of the magnetized stellar interior. If the star initially
is uncharged (as is assumed in our simulations), a surface charge of
opposite sign compensates the net charge of the interior of the star
(see \citealp*{MichelLi99} for thorough discussion). This surface
charge, together with the induced quadrupolar surface charge from
corotation, can leave the star and be redistributed throughout the
magnetosphere when conductivity is turned on, or when the work
function of the surface is low. Such redistribution lowers the maximum
available potential drop.

The available monopolar surface charge varies with inclination as
$\cos\alpha$ \citep{MichelLi99}, so the potential drop associated with
it disappears for orthogonal rotators. Hence, the value $V_{\rm
drop}\sim 0.2 V_0$ for the vacuum $90^\circ$ rotator in
Fig.~{\ref{potential}} represents the quadrupolar contribution to the
potential drop.  At conductivity $(\sigma/\Omega)^2 \sim 0.04$, the
maximal potential drop is close to the orthogonal vacuum rotator value
of $V_{\rm drop}\sim 0.2 V_0$ at all inclination angles.
Qualitatively, magnetospheres with finite conductivity
$(\sigma/\Omega)^2>0.04$ can effectively redistribute the surface
charges and maintain the linear relationship between the maximal
potential drop and the spin-down luminosity.  Below $(\sigma/\Omega)^2
= 0.04$, the surface charge redistribution is incomplete (the
monopolar surface charge is concentrated closer to the star in the
corotating steady-state solution) and the curves in
Fig.~{\ref{potential}} gain a tilt more reminiscent of the vacuum
solution.  The potential drops are not constant with inclination angle
at a given conductivity, and the linear relation (\ref{eq:spindown})
no longer holds.  We have explicitly verified that by the end of the
simulation these solutions reached steady-state in the corotating
frame.

Redistribution of surface charge in near-vacuum magnetospheres also
affects the spin-down power.  Looking back at Fig.~\ref{dipole} and
the piecewise linear relations (\ref{eq:spindownsigma}), we see that
the final 30\% of the spin-down transition between force-free and
vacuum rotators occurs for $(\sigma/\Omega)^2 < 0.04$. Despite the
fact that potential is no longer constant with angle for each value of
the conductivity in this regime, the transition in spin-down is still
continuous and smooth.

\section{Discussion}\label{sec:discussion}

We have presented a continuous one-parameter family of pulsar
magnetosphere solutions that span the range between the vacuum and
force-free limits.  Each solution of the family is characterized by
the value of the conductivity parameter, $\sigma/\Omega$, which is
related to the maximum potential drop that a test particle can
experience as it moves along field lines, $V_{\rm drop}$.  The zero
conductivity limit, $\sigma/\Omega\to0$, $V_{\rm drop}\to V_{\rm
drop,vac}$, yields the vacuum magnetosphere.  It shows a substantial
effective potential drop that makes up a significant fraction of the
rotation-induced potential difference between the pole and the equator
of the star ($20{-}60\%$ depending on the inclination).  The high
conductivity limit, $\sigma/\Omega\to\infty$, yields essentially an
ideal force-free magnetosphere. Unlike in vacuum, abundant
magnetospheric charges short out any potential differences along field
lines, leading to vanishing effective potential drop, $V_{\rm drop,ff}
\to 0$.  While in the vacuum case nearly all field lines, even those
that extend beyond the light cylinder, eventually return to the star
and are formally closed, in the ideal force-free case a fraction of
field lines open up and reach infinity. Our simulations show that
resistive high-$\sigma$ pulsars spin down at least $3$ times faster
than resistive low-$\sigma$ and vacuum pulsars (for the same value of
inclination), in agreement with earlier ideal force-free simulations
(S06).  Our resistive solutions bridge the gap between the force-free
and vacuum limits by having intermediate values of major
magnetospheric parameters, such as effective potential drop, $0<V_{\rm
drop}<V_{\rm drop,vac}$, the fraction of open field lines, and the
spin-down rate.

Before discussing possible observational implications of these
solutions, we note that our approach to modeling resistive pulsar
magnetospheres is quite simplistic: we use a form of Ohm's Law in
which we neglect several terms \citep{Meier04} and we assume constant
conductivity $\sigma/\Omega$ throughout all space.  The finite bulk
conductivity used in this work can be thought of as not due to
collisions, but due to the insufficient plasma supply in the
magnetosphere that cannot short out all the accelerating fields.  It
is possible that real magnetospheres have a lower (anomalous)
conductivity in the current layer and the current sheet than in the
rest of the magnetosphere. We might also expect a lower value of
conductivity on the open field lines than on the closed field lines.
Since our models do not explicitly model the plasma fluid
(\S\ref{sec:intro}), the plasma velocity component along the direction
of magnetic field is unconstrained, which necessitates the choice of a
frame in which to write the Ohm's Law.  We choose the so-called
minimum velocity frame, which has a number of attractive properties
(see \S\ref{sec:currentder}), yet this is not a covariant choice.
With these important caveats in mind, let us now for the sake of
argument take our model seriously as a description of pulsar
magnetosphere and consider the consequences.

Intermittent pulsars offer a unique test bed of pulsar theory (e.g.,
PSR B1931+24, \citealt{Kramer06}, and PSR J1832+0029,
\citealt{Lyne09}; see also \citealt{Wang07}, \citealt{Zhang07}, and
\citealt{Timokhin10}).  Such pulsars switch between an ``on'',
radio-loud, state in which they behave like normal radio pulsars, and
an ``off'', radio-quiet, state in which they produce no detectable
radio emission.  The two intermittent pulsars, for which published
data exist, have quite different duty cycles: PSR B1931+24, with a
period $P\approx0.8$~s, cycles through the ``on''--''off'' sequence of
states approximately once a month, whereas PSR J1832+0029, with a
period $P\approx0.5$~s, kept quiet for nearly two years.  The
spin-down rate for each of these pulsars is larger in the ``on'' state
than in the ``off'' state by a factor $f_{{\rm on}\to{\rm
off}}\simeq1.5$.  Such a substantial difference in spin-down rates
suggests that the pulsar magnetosphere undergoes a dramatic
reconfiguration as it transitions between the ``on'' and ``off''
states, yet such a transition was reported for PSR B1931+24 to take
place in just over $10$ pulsar periods.

\citet{Kramer06} propose that in the ``on'' state, plasma fills the
pulsar magnetosphere and supports (poorly understood) plasma processes
that produce radio emission.  In this picture, the pulsar transitions
to the ``off'' state when, due to some unknown trigger, magnetospheric
pair production shuts off: the remaining plasma leaks off the open
field lines and the pulsar goes radio-quiet.  In this scenario, we
expect the pulsar in the ``on'' state to spin down due to force-free
energy losses and in the ``off'' state due to vacuum dipole
losses. However, this picture is ruled out by observations: the
force-free spin-down rate is larger than the vacuum spin-down rate by
a factor, $f_{{\rm ff}\to{\rm vac}}\ge3$ (S06), that is clearly
incompatible with the observed value, $f_{{\rm on}\to{\rm
off}}\simeq1.5$ \citep[e.g.,][]{BN07}.

Our resistive magnetospheres provide a possible resolution to this
problem since they have intermediate spin-down rates between vacuum
and force-free.  Let us associate the ``on'' state with the force-free
magnetosphere and the ``off'' state with one of our intermediate
resistive magnetospheres.  According to equation (\ref{eq:spindown}),
for the ratio of spin-down powers in these two states to equal
$f_{{\rm on}\to{\rm off}}$, the potential drop has to be ${V_{\rm
drop}} = V_0{(f_{{\rm on}\to{\rm off}}-1)\left(0.9+1.1
\sin^2\alpha\right)}/{4.5 f_{{\rm on}\to{\rm off}}}$.  For $f_{{\rm
on}\to{\rm off}}\approx1.5$, the ratio $V_{\rm drop}/V_0$ increases
monotonically with increasing inclination angle, $\alpha$, from
$0.07$, for an aligned rotator ($\alpha=0^\circ$), to $0.15$, for an
orthogonal rotator ($\alpha=90^\circ$). Plugging in for $V_0$, using
the measured spin-down parameters of PSR B1931+24 (we associate the
observed spin-down rate in the ``on'' state with the spin-down rate of
the force-free magnetosphere), and assuming a neutron star mass,
$M_*=1.4 M_\odot$, we obtain $V_{\rm drop} =
1.5\times10^{16}(0.9+1.1\sin^2\alpha)^{1/2}~[{\rm V}]$.  This value
significantly exceeds the characteristic value of the polar cap
potential drop expected for pulsars \citep{GJ69, Arons09}, so the
pulsar should have no difficulty in creating pairs and shorting out
the large potential. The reason for this discrepancy is our
simplifying assumption that the conductivity is constant throughout
the magnetosphere.  This assumption causes the potential drop not just
across the polar cap but across the whole stellar surface.  In real
pulsars, we expect that the closed field line region is filled with
plasma and is highly conductive.  This plasma shorts out parallel
electric field in the closed field line region and screens the
potential drop everywhere except across the polar cap.  The potential
drop across the polar cap, $V_{\rm pc}$, is $\approx(R_*/R_{\rm LC})$
times the pole-to-equator potential drop, $V_0$, which gives
\begin{align}
V_{\rm drop} &\sim V_{\rm pc}\frac{(f_{{\rm on}\to{\rm
   off}}-1)}{4.5 f_{{\rm on}\to{\rm
   off}}}\\
&=3.8\times10^{12}~[{\rm V}], \notag
\end{align}
for PSR B1931+24, where we assumed $R_*=10$~km.

Addition of resistivity can, in principle, allow the detailed study of
pulsar spin-down and the determination of pulsar braking indices.  For
a spin-down law $\Omega = \Omega(t)$, the braking index is defined as
$ n \equiv {\Omega \ddot\Omega}/{\dot\Omega^2} $.  In order to measure
a braking index numerically, we run a series of simulations for
different values of angular frequency, $\Omega$, and map out the
dependence, $L(\Omega)$.  We find that $L(\Omega)\propto \Omega^4$ for
both purely force-free and vacuum solutions, which translates into a
value of the braking index, $n=3$, as we will see below.  This is
larger than the range of observed values, $n\simeq2.5{-}2.8$
\citep{Livingstone}.  One way to lower the braking index is to allow
the evolution of the extent of the magnetosphere (the radius of the
Y-point) to lag behind the outward expansion of the light cylinder due
to spin down \cite[e.g.,][]{ContopSpitkovsky06}. The extent of the
Y-point is controlled by the rate of reconnection at the edge of the
magnetosphere. We calculated the braking index in our simulations by
comparing the magnetosphere with constant conductivity spun up to
different periods, and we do not find significant deviations from
$n=3$.  This rapid spinup biases our answer, however, because the
Y-point extends to the light cylinder within one rotation period.  It
is not entirely clear that we would obtain the same braking index if
we allowed the pulsar period to increase during a single simulation as
the pulsar spins down and the light cylinder slowly recedes.  The
extent of the Y-point may depend on the past history of the pulsar.
To do this calculation, we need to increase the pulsar period several
times within a single simulation and measure the resulting spin-down
luminosity after each increase.  We must also ensure that the
reconnection at the Y-point is controlled by the resistivity we impose
and not by numerical effects.  A more accurate braking index
calculation would require much longer simulations, and possibly higher
spatial resolution near the Y-region.

Another possibility that can affect the braking index is the evolution
of conductivity with time. Such time-evolution translates into
evolution in pulsar dimensionless luminosity, $\ell = L/L_0$ (see
Fig.~\ref{dipole}), and leads to $n\neq3$:
\begin{equation}
  \label{eq:n}
  n = \frac{\Omega \ddot\Omega}{\dot\Omega^2} = 3 - 2 \frac{\mathrm
    d\ell/\mathrm dt}{\ell/\tau},
\end{equation}
where $\tau=P/2\dot P$ is the pulsar age, and we used the scaling,
$L_0\propto \Omega^4$. If dimensionless plasma conductivity does not
change in time, $\sigma/\Omega=$ constant, then $\mathrm \ell=$
constant and equation (\ref{eq:n}) gives $n = 3$.  Purely force-free and
purely vacuum pulsars, discussed above, fall into this category.  As a
pulsar ages, it is natural to expect that its plasma supply depletes
and the pulsar becomes progressively more vacuum-like: in other words,
both $\sigma/\Omega$ and $\ell$ decrease in time (see
Fig.~\ref{dipole}).  For such pulsars, $\mathrm d\ell/\mathrm dt<0$,
and equation (\ref{eq:n}) gives $n > 3$.  Of course, the converse is also
true, and the braking index lower than $3$ would result if a pulsar
becomes progressively more force-free--like, i.e., if pulsar
dimensionless luminosity increases in time, $\mathrm d\ell/\mathrm
dt>0$. Presently, however, it is not clear why such behavior would be
physically expected. Thus, the addition of bulk resistivity does not seem 
to easily solve the braking index puzzle, and a more elaborate 
explanation is still required. 

Our resistive magnetospheres naturally lend themselves to modeling
the sub-pulse drift phenomena observed in a number of pulsars
\cite[see
e.g.,][]{Askegar01,DeshpandeRankin,Rankin03,ContopSpitkovsky06}.  The
following discussion takes place in the corotating frame, where we
define electromagnetic fields by equations (\ref{eq:corotEfield}) and
(\ref{eq:corotBfield}).  In this frame there is transverse particle
drift with velocity of order $v_{\rm drift}\sim \vec{E}_{\perp}\times
\vec{B}/B^2$.  This drift velocity can be thought of as the minimal
velocity of particles, ignoring motion along the magnetic field lines.
The transverse electric field $\vec{E}_{\perp}$ reverses sign through
the axis of maximum potential drop, and so there is rotation of field
lines about this axis.  In general, this axis is not in the direction
of the magnetic moment, and in fact does not even need to be a
straight line.  If we consider a simplified picture of the
magnetosphere in which there is no parallel electric field in the
closed field line region, the potential drop across open field lines
is given roughly by the longitudinal potential drop along field lines.
The differential rotation can then be roughly related to the
longitudinal potential drop by
\begin{equation}
     \Delta\Omega = \Delta V_{\rm drop}/(B_p R_\perp^2) = \Delta V_{\rm drop} / \Psi_{\rm cap}.
     \label{eq:domegapc}
  \end{equation}  

In the laboratory frame, the field lines are undergoing rotation about
the pulsar spin axis (with the angular frequency $\Omega$), and in
addition they are rotating differentially around the maximum potential
axis in a retrograde sense.  The differential rotation of plasma can
explain sub-pulse emission features that drift with respect to the
otherwise periodic light curve of the pulsar.  Our result is in
contrast to the standard picture in which the plasma precesses about
the magnetic axis \citep{rudermansutherland75}.  The magnetic
colatitude parameter in the cartographic transformation of,
e.g. \cite{DeshpandeRankin}, must then be modified accordingly.

\begin{figure}[t]
\centering \includegraphics[scale=.9]{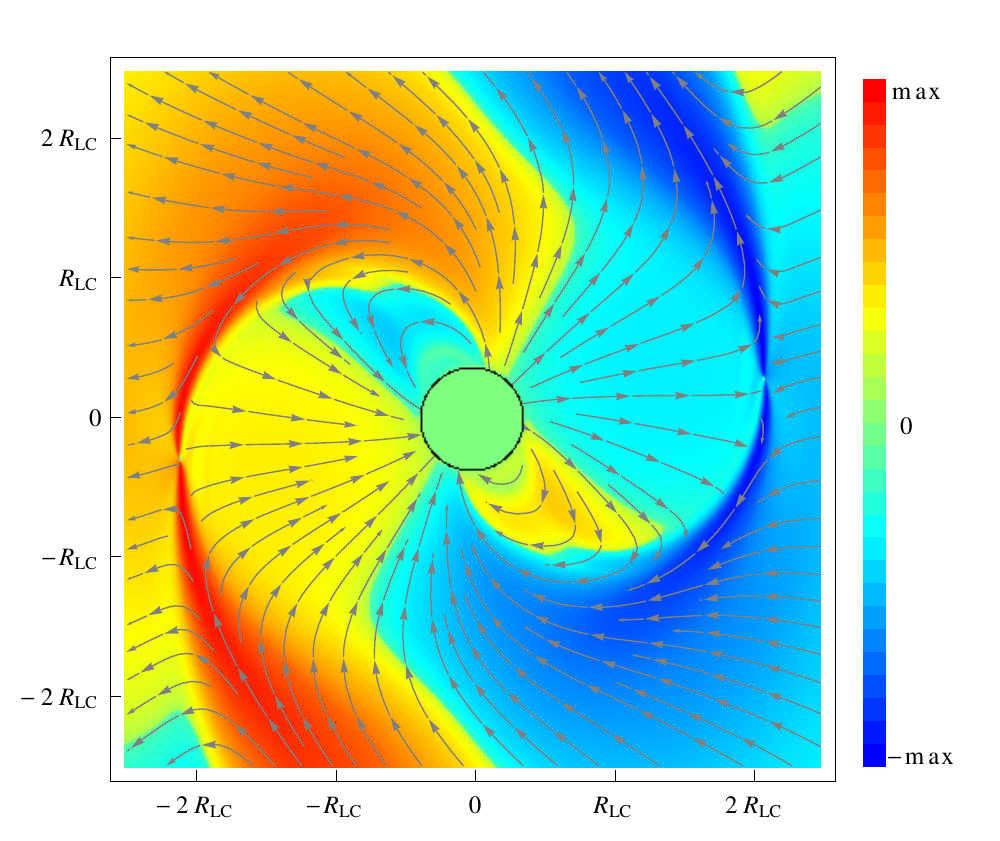}
\caption{Magnetic field lines in the $\vec{\mu}-\vec{\Omega}$ plane
for a $60^{\circ}$ inclined dipole at conductivity
$(\sigma/\Omega)^2=4$.  Color represents the out-of-plane drift
velocity in the corotating frame, with blue color representing the out
of the page direction.  There is retrograde differential rotation
about the axis with maximum potential drop, situated roughly halfway
between the rotation and magnetic axes when inside the light cylinder,
as well as about the boundary of the closed field line region.  The
maximum value on the color table corresponds to a drift velocity of
$0.8c$.}\label{velocity}
\end{figure}

To better visualize the differential rotation in the context of our
resistive magnetospheres, consider the full drift velocity
\begin{equation}
v_{\rm drift}=\frac{\vec{E}\times\vec{B}}{B^2+E_0^2}c,
\end{equation}
again in the corotating frame.  Fig.~\ref{velocity} shows magnetic
field lines in the $\vec{\mu}-\vec{\Omega}$ plane for a $60^{\circ}$
inclined dipole at conductivity $(\sigma/\Omega)^2=4$.  Color
indicates out-of-plane drift velocity, with red (blue) representing
velocity into (out of) the page.  The velocity has been rescaled by
raising its magnitude to the $1/2$ power.  The axis of maximum
potential extends from the star roughly halfway between the rotation
and magnetic axes.  The differential rotation of plasma about this
axis is denoted by the transition from red to blue across the axis,
indicating a reversal in the sign of the out-of-plane drift velocity.
Note that the axis bends off in the direction of open field lines
beyond the light cylinder.  These corotating frame drift velocities
can be quite substantial.  The maximum illustrated drift velocity
occurs near the current sheets and corresponds to a velocity of
$0.8c$.  The differential rotation is much stronger than that implied
by sub-pulse drift, but we expect the qualitative features of
differential rotation to persist in solutions with more realistic open
field line potential drops.  At lower conductivities the magnitude of
differential rotation about the axis of maximum potential drop
increases, and the conductivity is a parameter that can be tuned to
attempt to match observed sub-pulse emission features.

Incidentally, the bundle of field lines surrounding the axis of the
maximum potential corresponds to the field lines that carry the
largest distributed current density on the polar cap (see Fig. 4 in
\citealt{BS10}). If the strength of the current is associated with
radio emission, then the core emission would have a centroid that is
offset from the magnetic pole. The offset could be as large as half of
the polar cap radius. This would introduce potentially significant
modifications to the polarization sweep of the core radio emission and
cause deviations from the expected S-curve of \cite{RadCooke69}. For a
number of pulsars, it would imply differences in the inclination and
viewing angles inferred from the shape of the polarization sweep.

We conclude with prospects for future research.  We will generalize
our assumption that $\sigma/\Omega$ is constant in the
magnetosphere. In addition to having higher conductivity in the closed
zone, we will experiment with prescriptions for anomalous resistivity
in the current sheet.  These improvements will yield a more realistic
magnetospheric structure, that can be accurate enough for geometrical
modeling of gamma-ray light curves from pulsars. The gamma-ray pulse
formation is sensitive to the geometry of the magnetosphere, and, in
particular, to the field lines near the current sheet
\citep{BS10}. Deviations from the force-free geometry may be important
for modeling the light curves of older pulsars which require wider
gaps in the outer-gap models \citep{Watters09}.

AT acknowledges support by the Princeton Center for Theoretical
Science fellowship and by the National Science Foundation through
TeraGrid resources provided by NICS Kraken under grant number
TG-AST100040. AS is supported by NSF grant AST-0807381 and NASA grants
NNX09AT95G and NNX10A039G.  We thank the referee, Ioannis Contopoulos,
for comments that helped improve this paper.  The simulations
presented in this paper were performed on computational resources
supported by the PICSciE-OIT High Performance Computing Center and
Visualization Laboratory. This research used resources of the National
Energy Research Scientific Computing Center, which is supported by the
Office of Science of the US Department of Energy under contract
No. DE-AC02-05CH11231.

\appendix
\section{A. Resistive Current Derivation}
We derive here in full generality our resistive current prescription.
We start in the fluid frame, with charge $\rho_0$ and current flowing
along the electric field with magnitude $\sigma E_0$.  For
convenience, we pick the current and electric field to point along the
positive z axis.  The magnetic field can point along the positive or
negative z axis, depending on whether $B_0$ is positive or negative.
If we boost along the z axis with $\vec{\beta_1}=(0,0,\beta_z)$, the
current 4 vector in the boosted frame satisfies
\begin{equation}
\left[ \begin{array}{c} \rho'c \\ \\ \\ j'_z \end{array} \right] =
\begin{bmatrix} \gamma_z & & & -\beta_z \gamma_z\\ & & & \\ & & & \\ -\beta_z \gamma_z  & & & \gamma_z \end{bmatrix} \times \left[
\begin{array}{c} \rho_0c \\ 0 \\ 0 \\ \sigma E_0 \end{array} \right].
\end{equation}
Boosting again in the x direction transverse to the electric and
magnetic fields with $\vec{\beta_2}=(\beta_x,0,0)$,
\begin{equation}
\label{lorentzx}
\left[ \begin{array}{c} \rho c \\ j_x \\ \\ j_z \end{array} \right] =
\begin{bmatrix} \gamma_x & & & \\ -\beta_x \gamma_x & & & \\ & & & \\ & & & 1 \end{bmatrix} \times \left[
\begin{array}{c} \rho'c \\ 0 \\ 0 \\ j'_z \end{array} \right],
\end{equation}
we obtain for lab frame quantities the system of equations
\begin{equation}
\label{eq:rho}
\rho =\gamma_x \gamma_z \rho_0 -\gamma_x \beta_z \gamma_z \sigma E_0/c,
\end{equation}
\begin{equation}
j_x = -\beta_x c\rho,
\end{equation}
\begin{equation}
\label{eq:jz}
j_z = -\beta_z \gamma_z c\rho_0 + \gamma_z \sigma E_0.
\end{equation}
Rearranging equation (\ref{eq:rho}),
\begin{equation}
\rho_0=\frac{\rho}{\gamma_x \gamma_z}+\beta_z \sigma E_0/c,
\label{rho0}
\end{equation}
and plugging into equation (\ref{eq:jz}),
\begin{align}
j_z &= -\beta_z \gamma_z c(\frac{\rho}{\gamma_x \gamma_z}+\beta_z \sigma
E_0/c) + \gamma_z \sigma E_0 \nonumber \\
    &= -\frac{\beta_z}{\gamma_x}c\rho+\frac{1}{\gamma_z}\sigma E_0.
\label{jzgeneral}
\end{align}

We next determine the magnitude of the perpendicular boost in the x
direction.  The electromagnetic fields are invariant under the
boost $\vec{\beta_1}$.  After applying the boost
$\vec{\beta_2}$, the lab frame electromagnetic fields become
\begin{align}
E_x &=0 & E_y &=\gamma_x(E'_y-\beta_x B'_z)=-\gamma_x\beta_x B_0 & E_z &= \gamma_x(E'_z+\beta_x B'_y)=\gamma_x E_0 \nonumber\\
B_x &=0 & B_y &=\gamma_x(B'_y+\beta_x E'_z)=\gamma_x\beta_x E_0 & B_z &= \gamma_x(B'_z+\beta_x E'_y)=\gamma_x B_0.
\end{align}
Hence
\begin{equation}
B^2=\frac{\beta^2_xE^2_0+B^2_0}{1-\beta^2_x},
\end{equation}
\begin{equation}
\beta^2_x=\frac{B^2-B^2_0}{B^2+E^2_0},
\label{betax}
\end{equation}
and
\begin{equation}
\gamma^2_x=\frac{B^2+E^2_0}{B^2_0+E^2_0}.
\end{equation}
The lab frame x direction corresponds to the $-\vec E\times\vec B$
direction.  The y and z coordinates are rotated with respect to the
$\vec B$ and $\vec E_{\perp}$ directions.  We let $\theta$ denote the
angle between the z axis and the laboratory frame magnetic field and
assume without loss of generality that $\beta_x > 0$.  The current in the
$\vec{E}\times\vec{B}$ direction is then
\begin{equation}
j_{\widehat{\vec{E}\times\vec{B}}}=\beta_x c\rho.
\end{equation}
Noting that
\begin{equation}
E^2_{\perp}=\vec{E}^2-(E_0B_0/B)^2=B^2+E^2_0-B^2_0-E^2_0B^2_0/B^2,
\end{equation}
\begin{equation}
\beta_x=\sqrt{\frac{B^2-B^2_0}{B^2+E^2_0}}=\frac{E_{\perp}B}{B^2+E^2_0}.
\label{eq:betax}
\end{equation}
The boost $-\vec{\beta_1}=(-\beta_x,0,0)$ is the slowest boost from
the lab frame that brings the electric and magnetic fields parallel to
one another, and $\beta_x$ can be thought of as a generalized drift
velocity.  From $\beta_x$, we immediately obtain the transverse current
\begin{equation}
j_{\widehat{\vec{E}\times\vec{B}}}=\frac{\rho E_{\perp}B}{B^2+E^2_0}c.
\end{equation}

The current in the direction of the magnetic field is
\begin{align}
j_{\widehat{\vec{B}}}&=j_z \cos \theta=(-\frac{\beta_z}{\gamma_x}\rho c+\frac{1}{\gamma_z}\sigma E_0)\frac{\gamma_x B_0}{B} \nonumber \\
&=(-\beta_z\rho c+\frac{\gamma_x}{\gamma_z}\sigma E_0)\frac{B_0B+E_0 \vec{E}\cdot\vec{B}/B}{B^2+E^2_0}.
\end{align}
The current in the direction $\vec{E_{\perp}}$ is
\begin{align}
j_{\widehat{\vec{E_{\perp}}}}&=j_z \sin \theta=(-\frac{\beta_z}{\gamma_x}\rho c+\frac{1}{\gamma_z}\sigma E_0)\frac{\gamma_x \beta_x E_0}{B} \nonumber \\
&=(-\beta_z\rho c+\frac{\gamma_x}{\gamma_z}\sigma E_0)\frac{E_0E_{\perp}}{B^2+E^2_0}.
\end{align}
Hence the laboratory frame current vector can be expressed in full generality as
\begin{equation}
\label{generalcurrentappendix}
\vec{j}=\frac{\rho c\vec{E}\times\vec{B}+(-\beta_z\rho c+\sqrt{\frac{B^2+E^2_0}{B^2_0+E^2_0}(1-\beta_z^2)}\sigma E_0)(B_0\vec{B}+E_0\vec{E})}{B^2+E^2_0}.
\end{equation}
In \S\ref{sec:currentder} $\beta_z$ has been replaced by $\beta_{||}$
to emphasize that it is the magnitude of the boost along the parallel
electric and magnetic fields to the fluid rest frame.

\section{B. Numerical effects on spin-down power}
In principle we should be measuring spin-down values at the stellar
surface, but there is uncertainty in the stellar spin-down measurement
due to stair-stepping at the spherical inner boundary on our Cartesian
grid.  To avoid this issue, we measure spin-down at the light
cylinder.  Dissipation of Poynting flux inside the light cylinder
artificially suppresses our spin-down estimates measured at the light
cylinder, though.  We quantify these uncertainties by computing the
force-free spin-down values measured both on the star and at the light
cylinder.  The true force-free spin-down luminosity likely falls
within the shaded grey region in Fig.~\ref{dipole} bounded by the
stellar and light cylinder spin-down values.  There is also a
deviation from the vacuum Deutsch solution for our zero conductivity
solution (compare dashed line to $(\sigma/\Omega)^2=0$ line).  This
difference is due to our boundary conditions at the star.  The
smoothing of the fields across the stair-stepped boundary leads to
small charge bleed-off from the inside of the star to the outside. The
first term in equation~(\ref{current}) is then not identically zero
just outside the star even in the $\sigma=0$ case. This numerical
artifact does not affect the solutions with high conductivity when the
physical current exceeds the numerical smoothing current, but below
$(\sigma/\Omega)^2=4\times10^{-3}$ it can influence the field
structure and slightly modify the spin-down power.

\begin{figure}[t]
\centering \includegraphics[scale=.5]{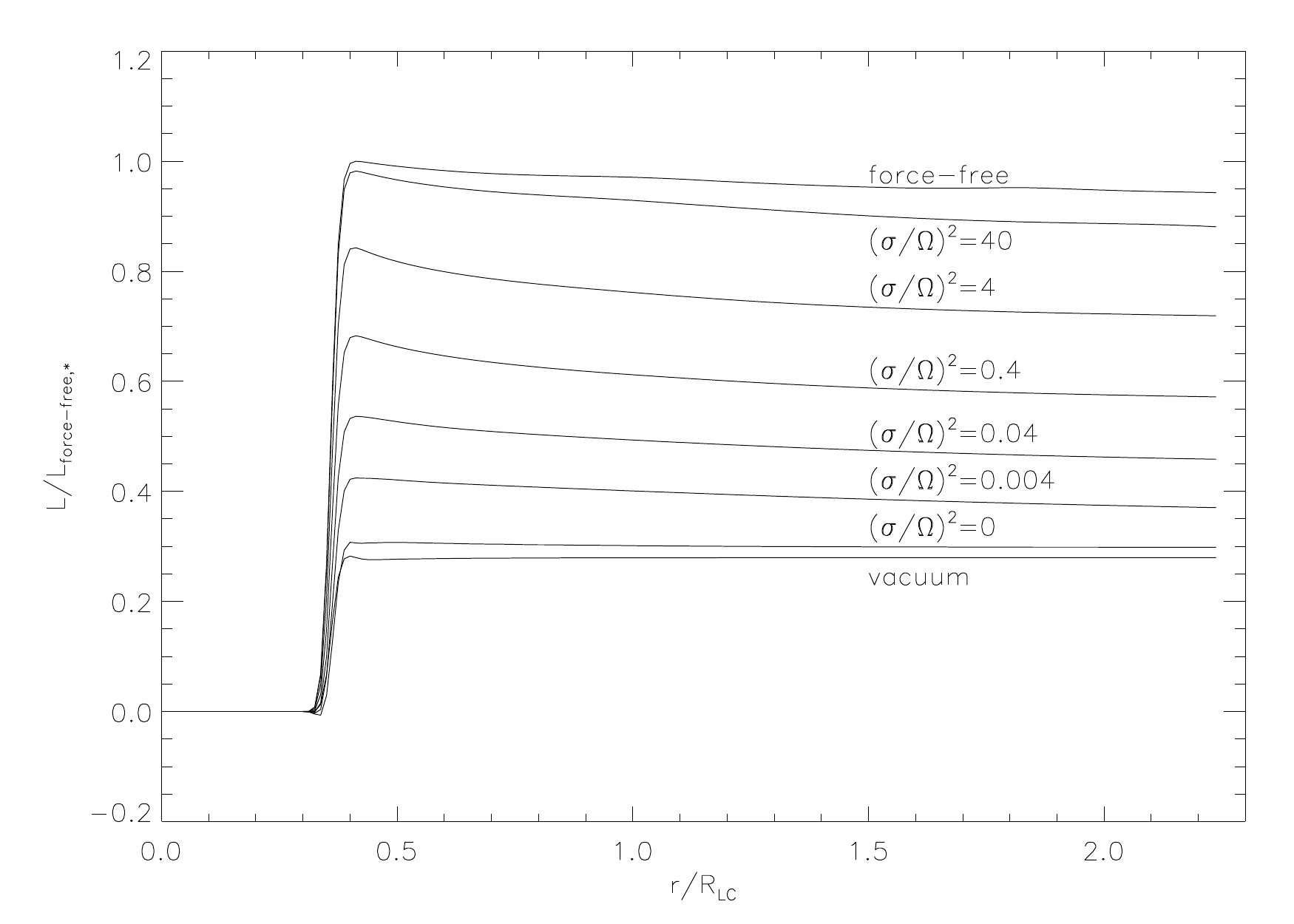}
\caption{Poynting flux with radius for force-free, a sequence of
resistive, and vacuum dipoles inclined at $\alpha=60^{\circ}$.  The
resistive Poynting flux values slope downwards with increasing radius,
reflecting magnetospheric dissipation.  Results are normalized to the
force-free spin-down value on the star.}\label{poynt}
\end{figure}

To better understand these numerical issues it is helpful to look at
the run of Poynting flux with radius.  In Fig.~\ref{poynt} we show the
Poynting flux integrated over spherical shells of varying radii for
force-free, vacuum and a range of resistive solutions with $\alpha =
60^\circ$.  The results are normalized to the force-free value at the
star, located at $R_*=0.375R_{LC}$, and we show the run of Poynting
flux with spherical radius out to $2.25R_{LC}$.  The vacuum and
$(\sigma/\Omega)^2=0$ curves are flat, indicating negligible
dissipation of Poynting flux with increasing radius, but there is an
offset between the curves that we attribute to our imperfect inner
boundary.  Force-free is in principle dissipationless and should have
the run of Poynting flux flat with radius.  As was argued above, our
method for cleaning parallel electric field in force-free simulations
is resistive and causes Poynting flux to slope downwards with
increasing radius.  Inside the light cylinder, the drop is due to
artificial volume $\vec{j}\cdot\vec{E}$ dissipation above the polar
caps.  This dissipation varies with magnetic inclination angle and is
responsible for the varying width of the gray error band with angle
for force-free solutions in Fig.~\ref{dipole}.  Outside the light
cylinder, the drop in Poynting flux is due to it disappearing into the
current sheets.

The resistive runs in Fig.~\ref{poynt} have Poynting flux dropping
with increasing radius, as we expect, indicating dissipation
throughout the magnetosphere.  Inside the light cylinder the
dissipation is primarily physical $\vec{j}\cdot\vec{E}$ dissipation
above the polar caps.  There is also some dissipation due to the
disappearance of Poynting flux into the current layer, though
significantly weaker in magnitude as compared to the volume
$\vec{j}\cdot\vec{E}$ dissipation.  Outside the light cylinder the
Poynting flux is dissipated in the current sheets.  The dissipation of
Poynting flux is strongest at intermediate conductivities,
$\sigma/\Omega\sim 1$.  At higher conductivities, the currents are
more ideal and dissipationless, and at lower conductivities,
dissipation is reduced because the conduction currents are
weaker. Volume dissipation introduced in our resistive simulations
implies that the light cylinder measurements of Poynting flux in
Fig.~\ref{dipole} underestimate the true spin-down power of the
resistive solutions, and should be treated as being within the error
bands similar to the gray band of the force-free solution.

\end{document}